\documentclass[prb,aps,twocolumn,epsfig,eqsecnum]{revtex4}

\usepackage{epsfig}
\newlength{\myL}

\newcommand{\beq}{\begin{equation}}
\newcommand{\eeq}{\end{equation}}
\newcommand{\bea}{\begin{eqnarray}}
\newcommand{\eea}{\end{eqnarray}}

\newcommand{\bB}{{\bf{B}}}
\newcommand{\bA}{{\bf{A}}}

\newcommand{\bE}{{\bf{E}}}
\newcommand{\bP}{{\bf{P}}}

\newcommand{\sxt}{\sigma^x_{\htau}}
\newcommand{\sxtt}{\tilde{\sigma}^x_{\htau}}
\newcommand{\hqdm}{H_{QDM}}
\newcommand{\br}{{\bf r}}
\newcommand{\bq}{{\bf q}}
\newcommand{\htau}{\hat{\tau}}

\newcommand{\wernercomment}[1]{}

\def\subsub#1{\noindent{\bf #1:}}

\def\state#1{\left|{#1}\right>}

\def\tit#1#2#3#4#5{{#1}{\bf #2}, #3 (#4)}

\def\npb{Nucl.\ Phys.\ B\ }

\def\prl{Phys.\ Rev.\ Lett.\ }
\def\pr{Phys.\ Rev.\ }
\def\prb{Phys.\ Rev.\ B\ }
\def\prd{Phys.\ Rev.\ D\ }

\def\sci{Science\ }

\def\jsp{J.\ Stat.\ Phys.\ }

\def\state#1{\left|{#1}\right>}

\def\bra#1{\left<{#1}\right|}
\def\braa#1{\left<{#1}\right.}

\def \x {{\bf x}}

\begin{document}


\title{Three dimensional resonating valence bond liquids and their excitations}

\author{R. Moessner$^1$ and S. L. Sondhi$^2$}

\affiliation{$^1$Laboratoire de Physique Th\'eorique de l'Ecole Normale
Sup\'erieure, CNRS-UMR8549, Paris, France}

\affiliation{$^2$Department of Physics, Princeton University,
Princeton, NJ 08544, USA}

\date{\today}

\begin{abstract}
We show that there are two types of RVB liquid 
phases present in three-dimensional
quantum dimer models, corresponding to the deconfining phases of $U(1)$ and
$Z_2$ gauge theories in $d=3+1$. The former is found on the bipartite
cubic lattice and is the generalization of  the critical point in the
square lattice quantum dimer model found originally by Rokhsar and Kivelson.
The latter exists on the non-bipartite face-centred cubic
lattice and generalizes the RVB phase found earlier by us on the triangular 
lattice. We discuss the excitation spectrum and the nature of the ordering
in both cases. Both phases exhibit gapped spinons. In the $U(1)$ case we 
find a collective, linearly dispersing, transverse excitation, which is the 
photon of the low energy Maxwell Lagrangian and we identify the ordering 
as quantum order in Wen's sense. In the $Z_2$ case all collective excitations
are gapped and, as in $d=2$, the low energy description of this topologically
ordered state is the purely topological BF action.
As a byproduct of this analysis, we unearth a
further gapless excitation, the pi0n, in the 
square lattice quantum dimer model at
its critical point.
\end{abstract}


\maketitle


\section{Introduction}
Fractionalised phases, i.e.\ phases where low energy excitations 
exhibit fractional quantum numbers, have played an important role in 
condensed
matter physics over the past two decades. Probably the two most
salient examples are provided by the quantum Hall effect, where the
charge of a quasiparticles can be a rational fraction of the
fundamental electronic charge,\cite{laughfqhe} and by Anderson's
suggestion of spin-charge separation being at the root of
high-temperature superconductivity.\cite{Anderson87}

This latter proposal has spawned a set of theories collectively
known as resonating valence bond (RVB) theories.\cite{ba,Rokhsar88} In
its short-range incarnation,\cite{Rokhsar88} RVB theory uses dimers 
as its
basic degree of freedom, which are cartoons for a pair of
adjacent electrons forming a singlet (`valence') bond. At zero doping,
when all sites are singly occupied, all electrons are involved in a
singlet bond with a neighbour. The Hilbert space of this theory is
thus given by the classical dimer coverings of the lattice.  Resonance
moves between dimer coverings provide a quantum dynamics for the dimers.

The original hope was that the resulting quantum dimer model would
have a liquid phase, the RVB liquid, hypothesized as an alternative to
Neel order in antiferromagnets by Anderson and Fazekas in the early
seventies.\cite{Fazekas74} This
state would exhibit fractionalized $S=1/2$ spinons and as a consequence
holes doped into it would undergo spin-charge separation and the resulting 
charged spinless quasiparticles (`holons') would Bose condense to form a
superconductor.

However, it turns out that the square lattice quantum dimer model
appropriate for the cuprates exhibits only solid phases with the
exception of one critical
point.\cite{Rokhsar88,subirfinitesize,fradkiv,levitov,leung} These
properties are now well-understood within the framework provided by
the height representation\cite{heights} for dimer
coverings.\cite{henleyjsp,msf}  By contrast, for dimer models
on non-bipartite lattices, the present authors have demonstrated that
an RVB liquid phase is possible.\cite{MStrirvb} The different outcomes
on bipartite and non-bipartite lattices can be rationalised, following
the ideas of Fradkin and Shenker on lattice gauge
theories,\cite{fradshen} via the absence (presence) of a deconfined
phase in $U(1)\ (Z_2)$ gauge theories in 2+1
dimensions.\cite{msf,subirvojta} We note that the RVB phase on
non-bipartite lattices is best characterized as a topologically
ordered phase\cite{wenniu} whose low energy theory is the purely
topological BF theory.\cite{hos}

This understanding of the situation in 2+1 dimensions has led us, in
the present work, to investigate the behaviour of quantum dimer models
in $d=3+1$, specifically on the bipartite cubic lattice and the
non-bipartite face-centred cubic lattice. We find that both models
exhibit RVB phases with deconfined spinons, which correspond to the
Coulomb phase of the $U(1)$ gauge theory (ordinary electromagnetism)
and the deconfined phase of the $Z_2$ gauge theory, respectively.  The
FCC lattice RVB liquid is topologically ordered. It has a gap to all
excitations and its ground state degeneracies and topological
interactions between spinons and vortex loops are encoded in the
topological 3+1 dimensional BF action.  By contrast, the cubic lattice
RVB phase supports a gapless, transverse collective mode. This can be
identified as the photon of the low energy theory, which is now a
non-topological Maxwell theory. Consequently, we identify the ordering
of this phase as quantum order in Wen's sense,\cite{wenquantum}
reserving the term topological order for cases where the low energy
theory is purely topological.  As a byproduct of this investigation,
we have also revisited the theory of the critical point of the square
lattice dimer model where we find, in addition to the resonons which
are the analogs of our photons in $d=2$, a further gapless excitation,
the pi0n, which signals the incipient crystalline order. We note that
our main results were announced in Ref.~\onlinecite{HKMS3ddimer}.

From the foregoing it follows that a deconfined phase in a valence
bond dominated regime is {\it easier} to achieve in $d=3$ than in
$d=2$, where its existence is ruled out for bipartite lattices. This
encouraging fact is, however, counterbalanced by the difficulty of
stabilising a valence-bond dominated phase in comparison to a Neel
phase, which becomes increasingly hard as the coordination of the
lattice grows. There is also the possibility of realizing dimer models
in other contexts -- e.g. the Santa Barbara group has shown that
multiple dimer models arise when frustrated Ising magnets are imbued
with a ring-exchange dynamics. Indeed in Ref.~\onlinecite{hermele03}
the reader can find a parallel discussion of a Colulomb phase in
multi-dimer models on the diamond and cubic lattices. 
Alternatively, dimer models can arise as Bose Mott insulators,
electronic Mott insulators at fractional fillings \cite{dhlsk}
and in mixed-valence systems on frustrated lattices.\cite{fuldemv}

We would be remiss if we did not note that quite generically
$U(1)$ gauge theories of Heisenberg magnets should be expected
to predict a Coulomb phase in $d=3$. Specifically,
a treatment of three-dimensional
quantum magnets in the bosonic large-$N$ theory\cite{arovasauer,readsachsun}
at a fixed (and small)
number of bosons per flavour, supplemented by $1/N$ corrections yields
results along the lines presented here for dimer
models. The dimer models themselves can be
obtained, at order $1/N$, in the large-$N$ limit taken with a fixed
boson number, and hence a vanishing number of bosons per
flavor.\cite{sachlect} It is a remarkable fact that these two very
different limits yield the same physics.  The equally 
remarkable fact that they also agree in $d=2$ was established previously.
It appears safe to conjecture that this agreement will hold
in higher dimensions as well.

Finally, we note that Wen has recently urged that the physical photon
be viewed as the consequence of a quantum ordered vacuum.\cite{wenlight}
The models
he has considered that give rise to an emergent photon have considerable
resemblence to the cubic dimer model considered in this paper. Indeed,
dimers and monomers have a natural string interpretation when superposed
on a reference configuration.

In the following, we first briefly summarise the known properties of
the quantum dimer model which are of relevance to us.  We
then describe the RVB liquids, first on the simple and second on the
face-centred cubic lattice. In both cases, we discuss in turn
correlations, topological properties and the excitation spectrum.

\section{Quantum Dimer Models: Generalities}
\label{sect:rk-qdm}

On a general lattice, allowed moves between two different hardcore dimer
configurations consist of interchanging alternately occupied and empty
links forming a closed loop, known as a resonance loop.  If we denote
the presence (absence) of a dimer by an Ising link variable,
$\sigma^x=+1 (-1)$, such a resonance move is generated by the kinetic
operator
\bea
\hat{T}_\circ=
\left(\prod_{i=1}^{n_l/2} \sigma^+_{2i-1}\sigma^-_{2i}\right) 
+ h.c.\ ,
\eea
where the loop, denoted pictorially by $\circ$, contains $n_l$ links,
and $\sigma^\pm$ are the raising and lowering operators for
to $\sigma^x$.
A corresponding projection operator, 
\bea
\hat{P}_\circ\equiv\hat{T}_\circ^2
\eea
determines whether dimers are arranged on a given loop so as to be
able to carry out a resonance move; such loops we call flippable
loops.

The quantum dimer Hamiltonian proposed by Rokhsar and Kivelson
considers only the shortest possible resonance loops. It contains two
terms, each involving one of these operators. The first carries out
resonance moves with a kinetic energy $t_\circ$, and the second exacts
an energy cost $v_\circ$ for a flippable loop:\footnote{When 
thinking of a dimer as a singlet bond, 
the sign of $t_\circ$ cannot be chosen freely. Whereas for the cubic
lattice, one can choose all $t_\circ>0$, this is not possible for the
FCC lattice. Moreover, on the FCC lattice, the different singlet
states are not linearly independent for $SU(2)$ spins. Interesting
things can happen as a result [see for example G. Misguich, D.  Serban
and V. Pasquier, cond-mat/0302152]. Since here we are interested in
quantum dimer models irrespective of connections to $SU(2)$ spins, we
do  not address these issues further and set $t_\circ>0$ uniformly.}

\bea
\hqdm&=&\sum_\circ -t_\circ \hat{T}_\circ+
v_\circ\hat{V}_\circ
\eea

\begin{figure}
\begin{center}
\epsfig{file=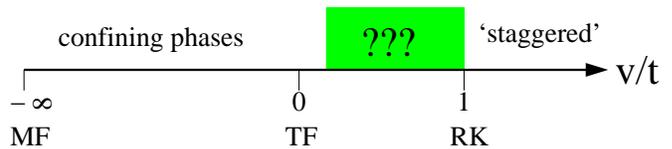,width=\columnwidth}
\end{center}
\caption{
Master phase diagram for Rokhsar Kivelon quantum dimer models.
Special points are highlighted: the Rokhsar-Kivelson (RK) point, the
transverse field (TF) point, and the maximally flippable (MF) point.}
\label{fig:masterRK}
\end{figure}

Some features are common to the phase diagrams of quantum dimer
models on most
lattices. We have depicted a generic phase diagram in
Fig.~\ref{fig:masterRK}. For $v/t>1$, it is favourable not to have any
possible flippable loops; configurations
satisfying this requirements can usually be found, and the resulting
phase is called the staggered phase, a name borrowed from the
appearance of one of the inert square lattice configurations.  For
$v/t=-\infty$, by contrast, one wants to maximise the number of
flippable loops. In both cases, there are no quantum
flucutations. These phases are confining, and confinement
typically exists for most values of $v/t$, with the possibility of
transitions between different confining phases.

From the point of view of deconfined, liquid\footnote{In the lattices
discussed in this paper the two terms are equivalent. On more
complicated lattices, such as the Fisher lattice discussed in
[R. Moessner and S. L. Sondhi,cond-mat/0212363], 
confining phases do not break
lattice symmetries and hence are indistinguishable from liquids from
the viewpoint of translational order.}  phases, the region $v/t\alt1$
is the most interesting, because the balance between kinetic term and
potential term overly favours neither high or low densities of
resonance loops. The ground state wavefunction can then be spread
moderately uniformly over most of the classical configuration space;
if the ensemble of classical configurations is sufficiently
disordered, this can then lead to an RVB liquid phase.

This statement can be made precise at the Rokhsar-Kivelson
(RK) point $v/t=1$. Define
a sector as the set of dimer coverings connected by $\hat{T}_\circ$.
Then the equal amplitude superpositions of dimer coverings
in each sector is a ground state at the RK point. Hence
equal time correlations can be computed as correlations of the 
uniform fugacity classical dimer model on the same lattice.\cite{Rokhsar88} 
These can often be obtained analytically -- exactly (especially in $d\leq2$) or
at least asymptotically -- or from Monte Carlo simulations. 

Consequently, advances in determining classical correlations give
valuable information on the quantum model. Indeed, it was the solution
of the classical dimer model on the triangular lattice with its
short-ranged correlations that allowed the triangular RVB liquid to
be discovered.\cite{MStrirvb} 

Recently, Huse, Krauth and the present authors have discussed the
classical dimer correlations on the cubic and FCC lattices.\cite{HKMS3ddimer}
In the next section we will build on this work to establish the
presence and nature of RVB liquid phases on these two lattices.

\section{Cubic lattice: an RVB U(1) liquid}
\label{sect:cub}

The general structure of the phase diagram on the cubic lattice is
indeed as sketched in Fig~\ref{fig:masterRK}.  For $v/t>1$, the ground
states are the non-flippable (`staggered') configurations, such as the
ones obtained by appropriately stacking the two-dimensional staggered
configurations.  For $v/t$ large and negative, there are six maximally
flippable columnar configurations, again obtained from stacking planar
columnar configurations. Apart from these two crystalline phases we
have been able to establish the existence of a liquid, RVB phase for
$v/t\alt1$.  This is a ``Coulomb'' phase, characterised by
algebraically decaying dimer (``magnetic field'') correlations and
whose spectrum contains a linearly dispersing transverse collective
excitation (``photon''), a gapped topological defect (``electric
monopole'') in addition to Coulombically interacting spinons
(``magnetic charges''). We now sketch the derivation of these results.

\subsection{RK Point}

We begin with the properties of the RK point, $v/t=1$. The classical
dimer model on the cubic lattice was studied in
Ref.~\onlinecite{HKMS3ddimer}. It was shown there that a useful
parametrization of the dimer configurations is in the language of
solenoidal magnetic fields.  Briefly, on the bipartite cubic lattice,
on can think of a dimer as a magnetic flux, $\bB$ of strength
$5/6=1-1/z$ pointing from sublattice A to sublattice B; here, $z$ is
the coordination of the lattice. Identifying an unoccupied bond with
flux $-1/6=-1/z$, one has $\nabla\cdot \bB=0$ -- magnetic charges are
excluded. 

Upon local coarse-graining, analogous to what is done for
models with height representations in two-dimensions, long wavelength
configurations acquire an entropic weight quadratic in $\bB$ and this
allows computation of the long distance correlations of the classical
model. These are purely dipolar,
\beq
\langle B_i(\x) B_j(0) \rangle = {1 \over K} {3 x_i x_j - r^2 \delta_{ij}
\over r^5}
\label{eq:classcorr}
\eeq
where $K$ can be determined via Monte Carlo
simulations.\cite{HKMS3ddimer} The language of magnetic fields also
allows a characterization of the sectors.  For a cube with periodic
boundary conditions, the flux through any surface that wraps around it
is invariant under local dimer moves and is invariant under lattice
translations of the surface.  In particular, if we let $\Sigma_i$ be
planes perpendicular to the cubic unit vectors ${\bf e}_i$, the fluxes
$\Phi^B_i$ through them are the invariants that characterize a given
sector of the dimer model. This flux is also known as a $U(1)$ winding
number.\footnote{The definition provided here works for a system with
even sidelengths. More generally, one has to resort to the
slightly more involved transition graph construction described in
Ref.~\onlinecite{Rokhsar88}} For an $L^3$ cube these range in
magnitude from 0 to $L^2/2$. The form of the correlations given above
is the average over all sectors, but also holds individually for all
sectors with $\Phi^B = |\sum_i \Phi_i^B {\bf e}_i|
\sim O(\sqrt{L})$,
which dominate the sum at large $L$. Finally, we note that
the ``staggered'' configurations maximize  $\Phi^B$, or more
accurately, the average magnetic field strength.

At the RK point equal amplitude superpositions in all magnetic flux 
sectors constitute degenerate ground states. As we will see below,
for $v/t \alt 1$ this degeneracy is lifted and only the sectors 
with  $\Phi^B \sim o(\sqrt{L})$ will be important in the low
energy analysis---so we will not need more classical information
than given in Eq.~\ref{eq:classcorr} above. 
For this reason it is sufficient
to focus on the zero flux sector at the RK point for now.

As the exact ground state wavefunction is known, we can use the
single mode approximation (SMA) for dimer models, pioneered
by RK in their original paper \cite{Rokhsar88} to get a good 
description of the excitation spectrum. We remind the reader
that the SMA will provide a rigorous upper bound on the 
excited state energy at each momentum. As RK's explanation was
rather compressed and their treatment incomplete in one important
respect in the $d=2$ case, we will provide a fuller account
between this section and the appendices.

\subsub{Single Mode Approximation} Let $\state{0}$ denote 
the ground state of the dimer model under
consideration. 
As above, we define $\sxt(\br)$ as the Pauli spin
operator the eigenvalues $\pm1$ of which correspond to presence and
absence of a dimer on the link at location $\br$. Here, we have added
the `direction' vector $\htau$ for clarity to distinguish between
dimers pointing in the different possible directions.

Next, we define the Fourier transform of the dimer density operator, 
\bea
\sxtt(\bq)\equiv\sum_\br\sxt(\br)\exp(i \bq\cdot\br)\ .
\eea
Acting with this operator on the ground state yields a state
\bea
\state{\bq,\htau}\equiv\sxtt(\bq)\state{0}
\eea
with nonzero momentum $\bq$ so that  
\bea
\braa{0}\state{\bq,\htau}=0\ .
\label{eq:smaorth}
\eea

Provided that 
\bea
\braa{\bq,\htau}\state{\bq,\htau}\neq0\ ,
\label{eq:nonz}
\eea 
this excitation
has
a variational energy of
\bea
E(\bq,\htau)&\leq&
\frac{\bra{0}[\sxtt(-\bq),[\hqdm,\sxtt(\bq)]]\state{0}}
{\bra{0}\sxtt(-\bq)\sxtt(\bq)\state{0}}\equiv\frac{f(\bq)}{s(\bq)},
\eea
where the numerator and denominator in the final line are
the oscillator strength -- $f(\bq)$ -- and structure factor -- $s(\bq)$. 
These
can be evaluated as ground state expectation values.

If it now so happens that the density $\sxtt(\bq_0)$ is a conserved
quantity, then $[\hqdm,\sxtt(\bq_0)]=0$, and thence
$E(\bq_0,\htau)=0$. From the orthogonality relation
(Eq.~\ref{eq:smaorth}) it then follows that this is a variational
upper bound on the energy of the `excited' state
$\state{\bq_0,\htau}$. 
Similarly, the behaviour of $f(\bq)$ near
$\bq_0$ can be used to determine a bound on the dispersion of the soft
excitations. Further, and again only if Eq.~\ref{eq:nonz} holds, a
smooth behaviour of $f(\bq)$ while $s(\bq)$ diverges, can
also be used to infer gapless excitations. Indeed, this is the classic
signature of incipient order.

RK identified a conserved quantity for the square lattice, which
generalises to the cubic case as the density of dimers pointing in a
given direction ($\htau=\hat{x}$, say) at wavevector
${\bq_0}=(q_x,\pi,\pi)$. This follows from the fact that the quantum
dynamics always creates and destroys pairs of neighbouring dimers.
This implies that the oscillator strength, $f(\bq_0)$, vanishes.  
In fact, at a wavevector $\bq_0+{\bf k}$, one finds 
\bea
f({\bf k})\propto ({\bf k}\times\htau)^2\ ,
\label{eq:fktr}
\eea
as outlined in Appendix~\ref{app:oscstr}. 
This form holds for all values
of $-\infty<v/t\leq1$.

This, however, does not lead to lines of zero energy excitations
because the structure factor also vanishes for all $\bq_0$ with
$q_x\neq\pi$ -- Eq.~\ref{eq:nonz} is not satisfied.  
In particular, at the RK point, for momentum
$\bq=(\pi,\pi,\pi)+{\bf k}$, it has the form of a
transverse projector, as one can see by Fourier transforming
Eq.~\ref{eq:classcorr}, 
\bea
s_{\hat{x}}({\bf k})
\propto \frac{k_y^2+k_z^2}{k^2}\equiv\frac{k_{\perp}^2}{k^2}\ .
\eea

This implies that only transverse excitations are generated by
$\sxtt$, a fact that traces back to the absence of monopoles,
$\nabla \cdot \bB =0$. The transverse nature of such excitations
has already been noted by Hastings.\cite{mattex}

It is important to note that these are the {\it only} gapless
excitations for the cubic lattice. By contrast, the square lattice
also exhibits gapless excitations near $(\pi,0)$ and $(0,\pi)$, where
there is a divergence of, respectively, $s_{\hat{x}}$ and
$s_{\hat{y}}$, signalling the immediate proximity of a crystalline
phase. This is consistent with the identification of the RK point as a
critical point terminating a crystalline phase. These excitations,
which we have christened pi0ns, are discussed further in
Appendix~\ref{sect:less}. The absence of pi0ns in $d=3$ signals that
the RK point does {\it not} abut a crystalline phase but instead
terminates an RVB phase. We will use this `soft pi0n theorem' again in
the analysis of the FCC lattice.

\subsection{Effective field theory and RVB phase}
The above RK point results can now be used to constrain a long
wavelength field theory valid in the proximity of this point.
From it, we will be able to deduce the existence and properties
of the neighboring RVB phase. 

To this end we follow Ref.~\onlinecite{HKMS3ddimer} and solve the
constraint $\nabla\cdot \bB=0$ by writing $\bB=\nabla\times\bA$ where
$\bA$ lives on the dual cubic lattice. As this representation comes
with a local gauge invariance, we will pick the gauge $\nabla \cdot
\bA =0$ to have a faithful representation of the dimer
states.\cite{fn-compact} If we now think of the time evolution
generated by the kinetic energy, this will involve flipping
plaquettes. It is straightforward to see that such a flip represents
an (essentially) spatially local change in $\bA$.  This fact and the RK
point properties allow us to write down the action,
\bea
{\cal S}=\int d^3x dt [ (\partial_t \bA)^2 - \rho_2 (\nabla \times
\bA)^2 - \rho_4(\nabla\times \nabla \times\bA)^2] \ .
\label{eq:actiongauge}
\eea
We can recognize Eq.~\ref{eq:actiongauge} as the restriction to $\nabla
\cdot \bA =0$ {\it and} $A_0=0$ of the manifestly
space-time gauge invariant action
\bea
{\cal S}=\int d^3x dt [\bE^2-\rho_2\bB^2-
\rho_4(\nabla\times\bB)^2] \ ,
\label{eq:actiongauge2}
\eea
where $\bE=\partial_t \bA - \nabla A_0$, is the electric field and 
$\bB = \nabla \times\bA$ as before.

For $\rho_2=0$, this action reproduces the equal time (classical)
dimer correlators at the RK point. It also yields a transverse photon
with $\omega \sim \rho_4 k^2$ in agreement with the SMA
dispersion. Further, we see that only gradients of $\bB$ enter the
Hamiltonian so that all flux sectors will yield degenerate ground
states, as is the case in the exact solution.  Finally, for $\rho_2 <
0$ the system will be driven to maximize the average value of
$|\bB|$---which is the case microscopically for $v/t >1$. All of this
confirms that Eq.~\ref{eq:actiongauge2} is the correct long wavelength
field theory with $\rho_2$ changing sign exactly at the RK
point. Readers familiar with this set of problems will recognize that
this is in precise correspondence with the treatment of the square
lattice dimer model\cite{henleyjsp,msf}
with the difference that in $d=2$ one solves the
constraint by writing $\bB = \nabla \times h$\cite{youngaxe} 
in terms of the scalar
height function.

Unlike in $d=2$, where the discreteness of the height field becomes
relevant when $\rho_2 >0$, in the present problem there are no relevant
operators beyond those listed in Eq.~\ref{eq:actiongauge2} when
$\rho_2 \ge 0$. Hence for $v/t <1$ the action is now that of the
standard Maxwell theory and gives rise to a linearly dispersing
transverse photon, $\omega \sim \rho_2 k$, at long wavelengths.

\subsub{Other excitations} The (gapped) spinons are represented 
by monomers which lead to
a local divergence $\nabla \cdot \bB = \pm 1$ on sublattice
A/B. Hence they carry a magnetic charge. As the Maxwell action
is invariant under the duality $\bE \leftrightarrow \bB$, 
we could also
recast the spinons as electrically charged particles. Either
way, they interact via an inverse square force law and are
therefore deconfined. At the RK point the vanishing of $\rho_2$
results in the force vanishing altogether.

In addition to the photon, there is a topological defect of
interest in the model with dimers alone. This is the electric
monopole, being the magnetic monopole that is present in lattice
$U(1)$ gauge theories in our dual labelling of the fields.
To see how it is constructed, consider solving the Gauss's law
constraint $\nabla \cdot \bE = 0$ that arises in the passage
from (\ref{eq:actiongauge2}) by writing $\bE = \nabla\times\bP$.
As $\bE$ lives on the dual lattice, $\bP$ lives again on the
direct lattice. [In quantizing the theory we take $\bP$ and
$\bB$ to be the conjugate variables on a link.] Following
the standard construction of the monopole, one can pick a classical 
$\bP$ such that
$\bE$ has the form ${\bf \hat r}/r^2$ at long distances from
a specified point of the dual lattice while a Dirac string of
plaquettes ensures that  $\nabla \cdot \bE = 0$ holds 
everywhere.\cite{polyakov}
The monopole state then can be written down in the dimer
representation as
\beq
e^{i \int d^3 x \bP \cdot \bB} |0\rangle
\eeq
where $ |0\rangle$ is again the ground state.
Requiring that the resonance energy cost along the string does
not diverge with system size quantizes the monopole flux.
The monopole generalizes the construction of the vortex at the
RK point in $d=2$\cite{2dmonopole} and is gapped as befits a deconfined
phase.  It was also noted in that analysis that the binding of 
vortices to
spinons could alter their statistics.\cite{goldhaber}
Likewise in $d=3$ the binding
of electric monopoles to bosonic spinons will lead to fermionic
dyons. The monopoles also interact via an inverse square force.

\subsection{Fractionalization and quantum order}
RVB liquid phases are fractionalized in that they support $S=1/2$
spinons.  In cases where there are no gapless excitations in the
spectrum, e.g.\ the RVB phase on the triangular lattice and the quantum
Hall liquids, the fractionalization is of a piece with a ground state
degeneracy sensitive to ground state topology and a topological
interaction between various gapped excitations---a complex of
properties named topological order by Wen.\cite{wenniu} 
In its pristine form,
topological order is unaccompanied by any symmetry breaking.  Most
succintly, the low energy theory for topologically ordered systems is
a topological field theory, the Chern-Simons theory for the quantum
Hall states\cite{wenniu} and the BF theory for the $Z_2$ RVB 
state.\cite{hos} We will see in
the next section that the FCC RVB phase is indeed topologically
ordered in this sense.

However, the situation with regard to the $U(1)$ phase discussed in
this section is different. The low energy theory is the Maxwell theory,
which is not topological. Concomitantly, the gapless photon makes the
distinction between ground states and excited states fuzzy. While at
the RK point there is an exact degeneracy between different magnetic
flux sectors, elsewhere in the RVB phase, this degeneracy is lifted by
the $\bB^2$ term in the action, the coefficient $\rho_2$ of which no
longer vanishes. The splitting thus generated is $L^3\times{\bf
\Phi^B}^2/L^4={\bf \Phi^B}^2/L$, which vanishes -- 
only algebraically -- in
the thermodynamic limit provided ${\bf
\Phi^B}=o(\sqrt{L})$.  For ${\bf
\Phi^B}=O(1)$, the
splitting has the same magnitude as the excitation energy of a photon;
this is no coincidence if one thinks of a photon as imposing a flux
modulated at the photon's wavelength, so that locally
a sector with a photon present looks like it belonged to a sector with
a slightly different ${\bf
\Phi^B}$.

As a matter
of principle, one can identify the minimum energy states in the
nonzero ${\bf \Phi^B}$ sectors as degenerate ground states while reserving
the term excitation for excited states in a given sector. Indeed,
at this level the symmetry between $\bE$ and $\bB$ implies that there
is another set of states with a net electric flux that also have an
energy of $O(1/L)$ and should be identified as members of the ground
state multiplet. These are the analogs, in $d=3$ of the height shift
mode discussed by Henley for the RK point\cite{henleyjsp} -- 
in $d=2$ the time derivative
of $h$ is the electric field. Nevertheless, such degeneracies would
be hard to detect against the background of excitations with similar
energies and so their utility appears to be limited.

Consequently, we follow Wen in identifying the $U(1)$ RVB liquid as
an instance of ``quantum order''\cite{wenquantum} 
wherein the masslessness of the
photon is attributed to the rigidity of the low energy
gauge structure.

\section{Face-centred cubic lattice: an RVB $Z_2$ liquid}
\label{sect:fcc}
The behavior of the FCC lattice quantum dimer model parallels that of
the triangular lattice problem in $d=2$. Like the triangular lattice,
this non-bipartite lattice does not admit a representation by
solenoidal fields. Instead, it exhibits topological sectors, not
connected under a local dynamics, that are labelled by an Ising
variable for each periodic direction on the lattice. This we call a
winding parity to distinguish it from the $U(1)$ winding numbers ${\bf
\Phi^B}$.  The winding parity is defined in an analogous manner, by
counting whether the number of dimers cutting the planes $\Sigma_i$ is
even or odd.

To the right of the RK point, there exists the standard non-flippable
phase (Fig.~\ref{fig:masterRK}); a non-flippable configuration can
again be obtained by appropriately stacking staggered square lattice
configurations. The identification of the phase for large negative
$v/t$ remains an open problem.

\subsection{RK Point and RVB Phase}

For the RK point on the 3-torus there are 8 exactly degenerate
ground states which exhibit exactly the same equal time
correlations in the thermodynamic limit. More generally, there 
are $2^p$ degenerate states for a system with $p$ independent 
non-contractible loops.
The classical analysis in Ref.~\onlinecite{HKMS3ddimer} 
has shown that the correlations decay
exponentially to zero, with an extremely short correlation length of
$\xi=0.4$ nearest neighbour distances. 

Turning now to collective excitations within the SMA, we note
that there {\it are} conserved quantities, the number of dimers in 
the $xy$ planes at wavevector $\bq_0=(0,0,\pi)$ and its symmetry
related counterparts. Nevertheless, the structure factor vanishes
as well at these wavevectors, quadratically by the analyticity in
momentum space guaranteed by the short range of the correlations
in real space.

To see this, consider grouping all the sites of a
finite FCC lattice with periodic boundary conditions in the
$z$-direction according to their $z$-coordinate. Let $N_z$ be the
number of sites with a given $z$-coordinate.
Now, for a given dimer configuration, define the number of dimers 
linking a pair of sites with neighbouring $z$-coordinates by
$\Delta_{z,z+1}$. Then the number of dimers in the $xy$ plane,
$n$, at coordinate  $z$ is simply
$2n(z)=N_z-\Delta_{z,z+1}-\Delta_{z-1,z}$. Therefore, for any
$k_z=q_z-\pi\neq0$,
\bea
\tilde{n}(q_z)&\sim&
-\sum_z \Delta_{z,z+1} \exp\left(i q_z z \right)
\left[ 2\exp\left(i q_z/2 \right)\cos q_z\right]\nonumber\\
&\sim& i (q_z-\pi) \tilde{\Delta}(q_z)\ ,
\eea
where the last term in parentheses was expanded in $k_z$; at
$q_z=\pi$, it vanishes linearly, yielding a quadratically vanishing
structure factor.

From the absence of any significant crystalline correlations or
gapless excitations at the RK point (no soft pi0ns) it is safe
to conclude that for $v/t <1$ there is a finite RVB phase with
liquid dimer correlations, an 8-fold ground state degeneracy on 
a 3-torus
and a gap to all excitations. Indeed the first and third pieces
are mutually consistent. The vanishing of the oscillator strength 
at  $\bq_0$ holds everywhere, while the
structure factor is also zero exactly at $\bq_0$. If the correlations
remain liquid, the structure factor will vanish analytically at
$\bq_0$ and hence the gap will persist. Evidently, matters will
be different when a solid phase is reached.

\subsection{Topological order}
The excitations of the FCC RVB phase include the ubiquitous
spinons and gapped vortex (vison) loops. The topological interaction
between them is described by the 3+1 dimensional BF action
which encodes the phase factor of $e^{i \pi n_l}$ for a
spinon trajectory that links $n_l$ times with the vortex
loop.\cite{hos} Quantization of the BF action on closed manifolds
recovers the ground state degeneracy discussed above and the
tunneling of the spinons and vortex lines lifts the
ground state degeneracy by an amount of $O(e^{-L})$ and
$O(e^{-L^2})$  for a system of linear dimension $L$. 

\section{Summary}
\label{sec:summary}

In this paper, we have established the existence of two types of RVB
liquids in $d=3$, which occur on the simple and face-centred cubic
lattices. As in $d=2$, the possible RVB phases in quantum dimer models
correspond to the known deconfined phases in compact gauge theories
and exhibit the same topological scaling limits. In addition to their
direct physical interest, they allow attractively simple illustrations
of concepts such as topological and quantum order.
 
Remarkably, the program outlined by Rokhsar and Kivelson in their
quest for the square lattice RVB liquid\cite{Rokhsar88} can be carried
through for the simple cubic lattice {\em in toto}. The ability to
perform calculations in a classical framework at the Rokhsar-Kivelson
point to access properties of both ground and excited states has been
invaluable for our study.
 
Finally, it will be interesting to see if such phases can be realised,
for example by destabilising Heisenberg Neel states on both lattices
using frustrating exchange interactions.
 
\section*{Acknowledgements}
We are grateful to Kirill Shtengel for asking stimulating
questions about the SMA analysis of quantum dimer models 
and specifically for sharing his own observations on the 
correlations of the dimer model on the square lattice. 
We also thank David Huse and Werner
Krauth for collaboration on closely related work.  RM would like to
thank Leon Balents, Matthew Fisher and Mike Hermele for several useful
discussions and SLS likewise thanks Vadim Oganesyan.
RM was in part supported by the Minist\`ere de la
Recherche et des Nouvelles Technologies with an ACI grant, and thanks
the KITP and the MPI-PKS Dresden for hospitality during part of this
work. This work was supported by the NSF with grants PHY99-07949
(KITP), DMR-9978074 and 0213706 (SLS), and by the David and Lucile
Packard Foundation (SLS).

\appendix
\section{Calculation of oscillator strength}
\label{app:oscstr}

To calculate the oscillator strength, $f({\bf
q})=[\sxtt(-\bq),[\hqdm,\sxtt(\bq)]]$, we need to consider the
commutation properties of the dimer density operator,
$\sxtt(\bq)\equiv\sum_\br\sxt(\br)\exp(i \bq\cdot\br)$ with the
quantum dimer Hamiltonian. To make contact with the original quantum
dimer paper, where the case of excitations near ${\bf Q}=(\pi,\pi)$
on the square lattice
were considered,\cite{Rokhsar88} we write $\bq={\bf Q}+{\bf k}$; more
generally, $2{\bf Q}$ needs to be a reciprocal lattice vector.

The dimer density operator commutes with the potential term of
$\hqdm$, so that we only consider its commutators with the
kinetic term. All that needs to be done here is to use
\bea
[\sigma^\pm,\sigma^x]=\mp 2 \sigma^\pm \ 
\eea
repeatedly. This form implies that the result in an isotropic phase
will always be the kinetic energy operator (and, at the RK point, the
expectation value of the loop flippability) times a function of $\bq$,
the details of which we are interested in. 

The contribution of a given square 
plaquette with dimers in the $\htau$ directions at locations ${\bf
R}$ and ${\bf R}^\prime={\bf R}+{\bf r}$ is given by
\bea
[\sxtt(-\bq)&,&[-t_\circ \hat{T}_\circ,\sxtt(\bq)]]
=\nonumber\\
&&8 t_\circ \hat{T}_\circ 
\exp(2i {\bf Q}\cdot{\bf R})
(1+\cos(({\bf Q+k})\cdot{\bf r}))\ .
\eea
Summing this expression over all plaquettes forces $2 {\bf Q}$ to be a
reciprocal lattice vector, and yields the expression proportional to
the kinetic energy operator.

The dependence on ${\bf k}$ and ${\bf Q}$ now follows
straightforwardly. For example, for the case of resonons on the square
lattice, ${\bf Q}=(\pi,\pi)$, and ${\bf r}=\pm\hat{y}$ for
$\htau=\hat{x}$. Thence, $f({\bf k})\propto 1-\cos k_y\sim k_y^2\sim
({\bf k}\times\htau)^2$, the original result of Rokhsar and
Kivelson.\cite{Rokhsar88} This form also obtains for ${\bf
Q}=(0,\pi)$, whereas near ${\bf Q}=(\pi,0)$,
$f({\bf k})\propto\cos^2(k_y/2)$ is nonzero everywhere.

The same analysis carries over directly to other lattices.  On the
cubic lattice, there are two transverse directions, and $f({\bf
k})\sim ({\bf k}\times\htau)^2$ continues to hold near ${\bf
Q}=(\pi,\pi,\pi)$.

For the hexagonal case, where an elementary resonance move involves a
triplet of dimers, one finds the resonon to occur near zero
wavevector, ${\bf Q}=0$, with the same form of $f({\bf k})\sim ({\bf
k}\times\htau)^2$.

On the triangular lattice, the same algebra yields a qualitatively
different result. If we consider the triangular lattice as a square
lattice decorated with diagonal bonds pointing from the botton left to
the top right corner of each square, the transverse projector form
for ${\bf Q}=(0,\pi)$
is replaced by $f({\bf k})\sim k_y^2+(k_y+k_x)^2$, which vanishes
quadratically as $k\rightarrow 0$ but is nonzero in all directions
away from $k=0$.

Finally, for the FCC lattice, where the conservation law involves two
different dimer directions, the algebra is more
complicated but one again finds a vanishing $f(k)$ as the wavevector
approaches ${\bf Q}=(0,0,\pi)$.

\section{SMA results in \lowercase{$d=2$}}
\label{sect:less}
\subsection{Resonons and pi0ns on the square lattice}
\label{subsect:squex}
For the bipartite square lattice, the dimer density $\sxtt(\bq)$\ is a
conserved quantity at $\bq_0(\htau,q_{\htau})={\hat{z}}\times\htau+
q_{\htau}\htau$ for any value of $q_{\htau}$.  Thus, $f(\bq_0)$
vanishes. As in the cubic lattice, this does not
lead to an  entire line of zero-energy excitations along the
Brioullin zone edge. Rather, the resonon only exists as a transverse
excitation near $(\pi,\pi)$.

To demonstrate this, we have explicitly determined the structure
factor for $\sxtt(\bq)$ and $\htau=\hat{x}$. The calculation of the
structure factor for two-dimensional models at the RK point is
straightforwardly achieved using the Fermionic path integrals
developed for the study classical dimer models. A detailed calculation
of correlations for the triangular lattice has been presented in
Ref.~\onlinecite{FMStridimer,ioselevich}.

\begin{figure}
\begin{center}
\epsfig{file=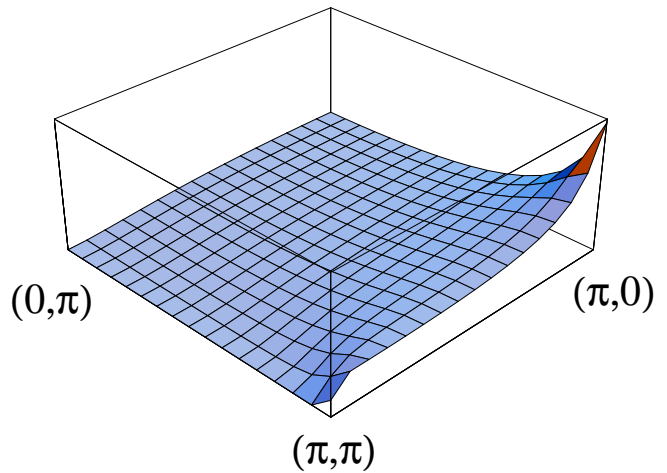,width=\columnwidth}
\end{center}
\caption{
Structure factor, $s_{\hat{x}}(\bq)$, for the square lattice.  The
values of $\bq$ are indicated on the plot. For $q_y\pi$, the structure
factor vanishes on the interval $q_x\in\ ]-\pi,\pi[$. It is positive
elsewhere, and at $\bq=(\pi,0)$, it diverges logarithmically.  }
\label{fig:sqstrfact}
\end{figure}

The result is plotted in Fig.~\ref{fig:sqstrfact}. The structure
factor vanishes everywhere on the line $(q_{\hat{x}},\pi)$ except for
at ${\bf Q}=(\pi,\pi)$. This implies that
$\braa{\bq,\htau}\state{\bq,\htau}=0$ on this line away from
$(\pi,\pi)$, so that it is not a candidate for a
gapless excitation.

In fact, the form of the structure factor near ${\bf Q}$ is
well-described by the transverse projector
\bea
s_{\hat{x}}(\bq)\sim\frac{k^2-({\bf k}\cdot\htau)^2}{k^2}
\sim\frac{k_y^2}{k^2},
\eea
where ${\bq}={\bf Q}+{\bf k}$.  Together with\cite{Rokhsar88} $f({\bf
k})\sim 1-\cos k_y\sim k_y^2=({\bf k}\times \htau)^2$, this implies
that there is only a single gapless excitation per dimer direction,
namely at ${\bf Q}=(\pi,\pi)$, which is of a transverse nature.

In Fig.~\ref{fig:sqstrfact} a peak in $s_{\hat{x}}(\bq)$ is visible at
$\bq_1=(\pi,0)$. Its scaling with system size can be determined easily
since the long-distance part of the correlations of the classical
dimer model in real space are known exactly:\cite{fisher63,henleyjsp}
\bea
C_{\hat{x}\hat{x}}(\br)\sim
(-1)^{x+y}
\frac{y^2-x^2}{r^4}
+(-1)^x\frac{1}{r^2}\  .
\label{eq:sqdimcorr}
\eea
This form implies that the peak height grows logarithmically with
system size, and equivalently that its height a distance ${\bf
k}=\bq-\bq_1$ from the centre decays (isotropically) as $-\log k$.

Near $\bq_1=(\pi,0)$, the oscillator strength depends on ${\bf k}$
simply as $f({\bf k})\sim\cos^2(k_y/2)$, which is a constant to
leading order. Therefore, $E(\bq_1,\hat{x})=0$, and there exist
further gapless excitations for each dimer direction, which -- due to their
location in reciprocal space -- we call pi0ns. The dispersion
obtained within the SMA is logarithmic, but more on that below.

We mention in passing that the same analysis carries over to the
hexagonal lattice, where the resonon exists at ${\bf Q}=(0,0)$ and the
now inappropriately named pi0n at $(4\pi/3,0)$.

This similarity is efficiently encoded in the language of
two-dimensional height models, analysed in detail by
Henley.\cite{henleyjsp} Both resonon and pi0n can be connected back to
slow modes in the height model at zero wavevector, the connection to
the dimer modes being via the gradient and vertex operators,
respectively. Indeed from the height analysis one learns\cite{ms-spinice}
that the structure factor at $\bq_1 + {\bf k}$ decays in imaginary
time as $e^{-k \sqrt{\tau}}$ which confirms that there are gapless
excitations near $\bq_1$ but also shows that the SMA does not do a 
very good job of constructing them.

\subsection{Absence of gapless modes on the triangular lattice}
\label{subsect:trisma}
The argument that the SMA mode for on the triangular lattice is gapped
at all momenta, despite the presence of a conservation law, is
completely analogous to the one presented above for the FCC
lattice. It again involves showing that the corresponding structure
factor vanishes at least quadratically.  The relevant momentum for
dimers pointing in the $\hat{x}$ direction is ${\bf Q}=(0,\pi)$.  A
direct computation of the structure factor using the results of
Ref.~\onlinecite{FMStridimer} yields the same result. This SMA result
was already noted in Ref.~\onlinecite{MStrirvb}.


\begin{thebibliography}{99}
\bibitem{fn-compact}
On the torus there is a non-trivial loop in the space
of dimer states that, stictly speaking, makes this mapping 
non-unique. However that can be ignored in what follows.
\end{thebibliography}
\end{document}